# A prototype station for ARIANNA: a detector for cosmic neutrinos

Lisa Gerhardt, Spencer Klein, and Thorsten Stezelberger
Nuclear Science Division, Lawrence Berkeley National Laboratory, Berkeley CA, 94720 USA

Steve Barwick, Kamlesh Dookayka, and Jordan Hanson
Department of Physics, University of California, Irvine, CA 92697, USA

Ryan Nichol
Department of Physics and Astronomy, University College London, United Kingdom

## Abstract

The Antarctic Ross Iceshelf Antenna Neutrino Array (ARIANNA) is a proposed detector for ultra-high energy astrophysical neutrinos. It will detect coherent radio Cherenkov emission from the particle showers produced by neutrinos with energies above about $10^{17}$ eV. ARIANNA will be built on the Ross Ice Shelf just off the coast of Antarctica, where it will eventually cover about 900 $km^2$ in surface area. There, the ice-water interface below the shelf reflects radio waves, giving ARIANNA sensitivity to downward going neutrinos and improving its sensitivity to horizontally incident neutrinos. ARIANNA detector stations will each contain 4-8 antennas which search for brief pulses of 50 MHz to 1 GHz radio emission from neutrino interactions.

We describe a prototype station for ARIANNA which was deployed in Moore's Bay on the Ross Ice Shelf in December 2009, discuss the design and deployment, and present some initial figures on performance. The ice shelf thickness was measured to be $572 \pm 6$ m at the deployment site.

## Introduction

Ultra-high energy (UHE) astrophysical neutrinos are a compelling target in high-energy particle astrophysics. Cosmic-ray protons with energies above about $4\times10^{19}$ eV can interact with ambient cosmic microwave background photons, forming a $\Delta^+$ resonance. The $\Delta^+$ decay starts a chain whose final products include neutrinos; these are known as GZK neutrinos [1].

Observation of a significant (100 event) sample of GZK neutrinos would provide a wealth of physics information on the origin and composition of UHE cosmic rays, and would also probe some important questions in particle and nuclear physics [2]. For example, a measurement of the neutrino-nucleon cross-section would provide information about low-x parton distributions in nuclei.

Despite much effort, GZK neutrinos remain as-yet unobserved; only now are current experiments beginning to set limits on their flux. These experiments look for optical, acoustic [3] or radio emission from neutrino showers. The current best limits come from the AMANDA optical Cherenkov detector at the South Pole [4], and from the ANITA balloon experiment [5]. ANITA circled Antarctica twice, in 2006-7 and 2008, at an altitude of 35,000 km, looking for radio waves emitted by neutrino interactions in the Antarctic ice; ANITA currently has the most restrictive limits at energies above about $10^{19}$ eV [5]. Other groups have looked for evidence of neutrino interactions in Greenland [6] or in the horizontal air showers in the atmosphere [7]. Other experiments have also looked for radio emission from neutrino interactions in the moon [8]. In these 'standoff' experiments, where the neutrino interaction target is separated from the detectors, the threshold is typically above $10^{19}$ or $10^{20}$ eV, enough only for the upper tail of GZK neutrinos.

Other experiments, such as ARIANNA [9], RICE [10] and the proposed ARA [11] experiments, at the South Pole, avoid this problem by co-locating their receivers in the detection volume, so they have lower threshold energies, of order $10^{17}$ eV. ARIANNA benefits from the strong radio reflectivity of the ice-water interface below the ice shelf; this interface reflects radio waves from downward-going neutrinos, greatly increasing ARIANNA's angular coverage [9]. Since UHE neutrinos are absorbed by the earth, this leads to a large increase in effective area.

## Radio emission from neutrino showers

Particle showers from neutrino interactions in Antarctic ice produce short duration (~ 1 ns) radio pulses via a process known as the Askaryan effect [12]. The pulse occurs because electromagnetic and hadronic showers produced in neutrino interactions (or by any other high-energy interaction) contain more electrons than positrons. The excess electrons are created late in the shower, when the average shower particle energy is near the critical energy, 79 MeV in ice; this is when the shower multiplicity is at its highest. Then, photons Compton scatter from atomic electrons, while shower positrons can annihilate on atomic electrons. These factors lead to a net negative charge; the magnitude of the excess is about 25 % of the total number of charged particles in the shower [13].

As these particles move faster than the speed of light in the medium, they emit Cherenkov radiation. At wavelengths that are large compared to the lateral spread of the shower, the Cherenkov amplitudes add coherently, and the Cherenkov radiation scales as the square of the charge excess, i.e. as the square of the neutrino energy. In ice, this coherence holds for radio waves with frequencies up to a few GHz. Near the maximum frequency, the radiation is narrowly concentrated near the Cherenkov angle; as the frequency decreases, the spread around the Cherenkov angle becomes wider. Accelerator studies have shown that calculations of the Askaryan effect are accurate [13].

## ARIANNA – the concept

GZK neutrinos travel great distances though the universe and arrive at the earth isotropically. Since the GZK neutrinos are energetic enough to be absorbed while travelling through the earth, most of the neutrinos that can be observed near the surface are either travelling near horizontally or in a downward direction.

ARIANNA observes the radio pulses generated by UHE neutrino interactions with oxygen and hydrogen nuclei in the ice of the Ross Ice Shelf. These interactions produce a shower of particles which in turn produce a radio pulse with a forward directed conical emission pattern; radiation is centered on the Cherenkov angle, about 41 degrees in ice. If the neutrino is travelling downward, the cone of radio waves will travel down to the saltwater-ice boundary (located 570 m beneath the snow surface), which acts like a mirror and reflects the pulse back to the top surface of the ice shelf. This reflection greatly increases the angular acceptance of ARIANNA and drives its design. ARIANNA consists of an array of autonomous stations that are deployed on the surface. Each station consists of 4-8 directional antennas. To maximize the sensitivity to reflected events, the antennas are oriented to point downward.

ARIANNA will be sited in Moores Bay on the Ross Ice Shelf ($78^0$ 44' 523" S, $165^0$ 02' 414"), which is about 110 km south of McMurdo station, the main US base in Antarctica. The stations will be installed over a 30 km x 30 km region. A high ridge, Minna Bluff, separates Moores Bay from McMurdo, so it blocks the vast majority of anthropogenic radio noise. Except for a few aircraft overflights a day during the Austral summer, we expect very small levels of impulsive radio noise at the ARIANNA site.

## The Prototype Detector

To test out the ARIANNA concept, and also to learn more about the site and the technology, we built a prototype detector which was deployed at the ARIANNA site for 1 year. The main goals were to study:

- Radio signal attenuation in the ice and reflection at the ice-seawater interface
- Radio backgrounds, especially anthropogenic noise
- Wind speed, with an eye to using a wind generator to power the station
- Temperature profiles over the year
- Performance of the prototype hardware in the Polar environment

## Detector Overview

Figure 1 shows a block diagram of the prototype detector. Four log-periodic dipole antennas are buried in the Ross Ice Shelf in Moore's bay. Each antenna feeds a low-noise amplifier, which itself feeds a switched capacitor array (SCA) analog-to-digital converter (ADC) and programmable trigger circuit. The SCAs are digitized and read out whenever the trigger fires. The standard trigger requires signals in at least two of the four antennas.

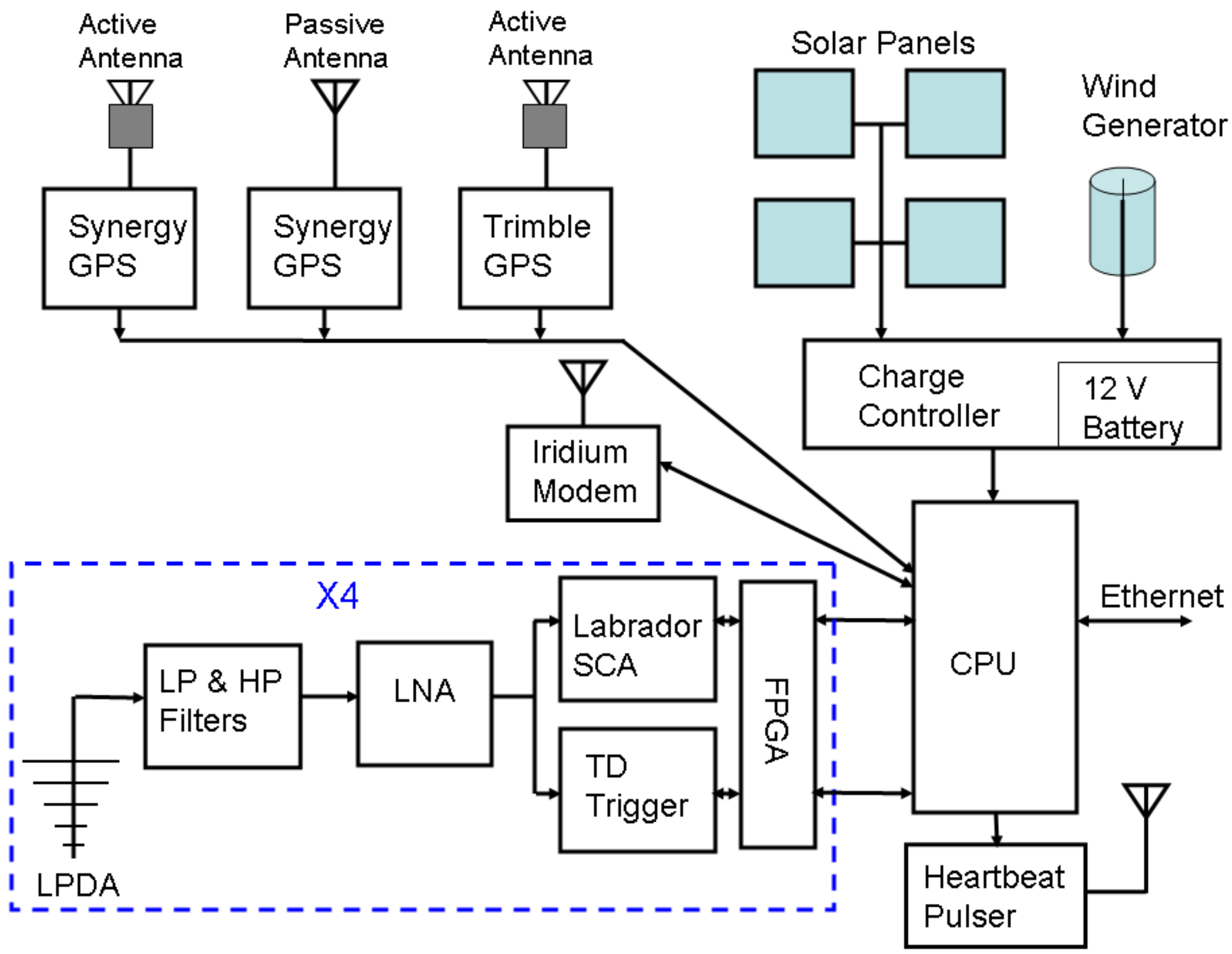

**Figure 1.** Block diagram of the ARIANNA prototype electronics. Each log periodic dipole antenna (LPDA) feeds low-pass (LP) and high-pass (HP) filters (in series), a low noise amplifier (LNA) which go to a tunnel diode (TD) trigger and the Labrador switched capacitor array (SCA) digitizer. The trigger and SCA are controlled by a field programmable gate array (FPGA). The 3 GPS receivers were for comparison purposes. The 'heartbeat' pulser transmits test pulses.

The system is controlled by a PC104plus linux-based computer with an 8 GByte flash disk. GPS receivers are used for accurate time-keeping. The station has 3 GPS receivers, so we can compare two different models; for one model we are comparing two versions, with active and passive antennas.

The station communicates via an Iridium satellite modem. During the summer, it also had a wired Ethernet connection which was connected to a wireless Bridge station. This station communicated with a repeater which was installed on Mt. Discovery, about 40 km away, which in turn connected to McMurdo Station. When fully operational, the station consumes about 25 Watts. During the summer, power is provided by four 30-Watt solar panels. In the winter, a Forgen 1000LT wind generator will provide some power. The power controller includes a gel battery to buffer the generators through periods of darkness (during the spring/fall) and/or low winds. The computer can turn off various pieces of the station (including data collection) to reduce the power consumption.

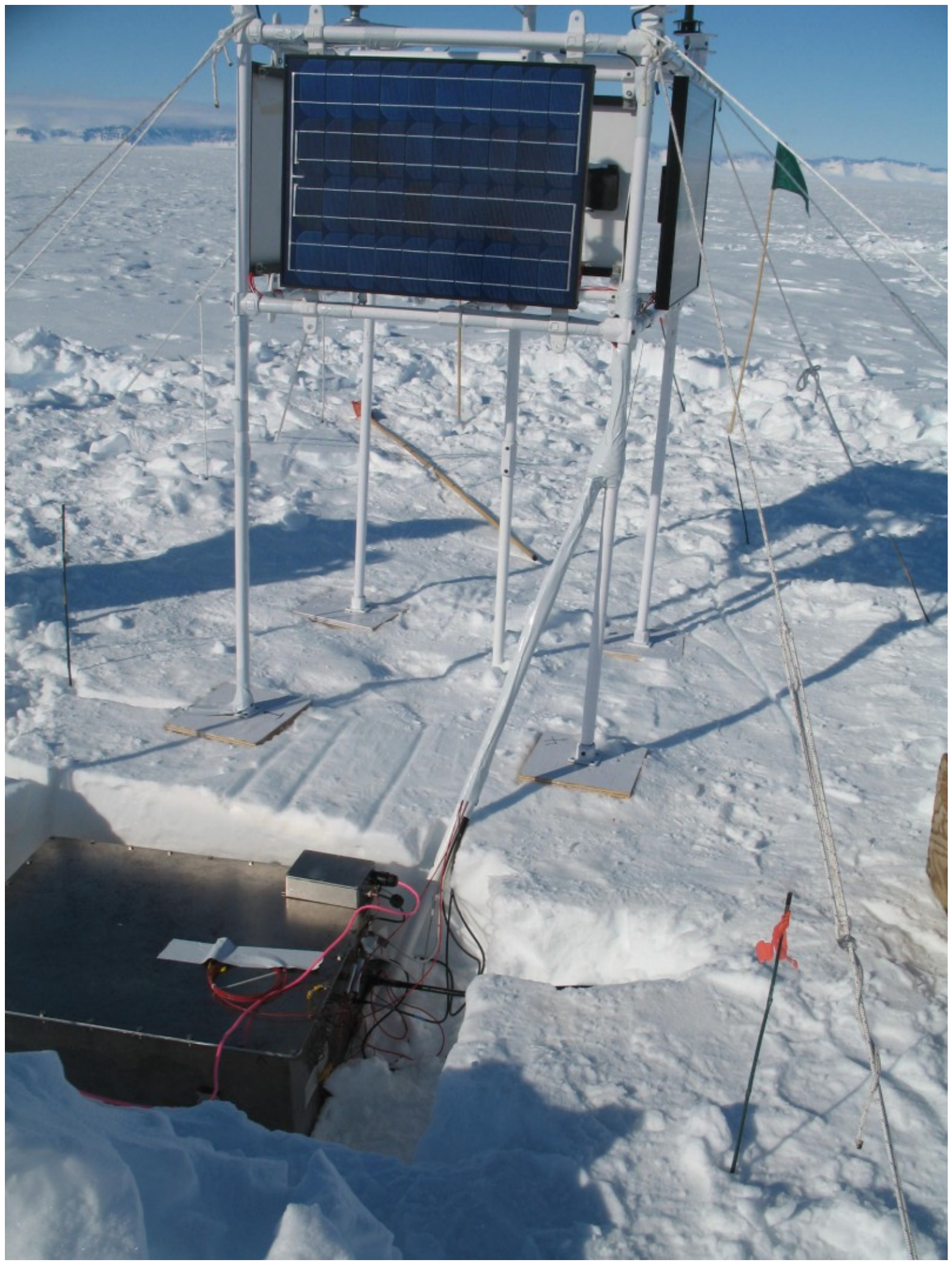

**Figure 2.** A photo of the ARIANNA station during installation, showing the tower structure and solar panels. The main electronics box is visible in the foreground, before being covered in snow. The small box with the pink cable is an Ethernet converter module. The plywood supports were later buried in snow. For scale, the four tower support legs form a square with sides about 96 cm long.

Most of the electronics are in a steel box which is buried with its top flush with the snow level. The power controller and battery are in a separate box, which was buried about a foot away. The antennas, four solar panels, a wind generator and an anemometer are mounted on a square tower structure shown in Fig. 2. The tower is constructed of aluminum pipe held together with cast aluminum fittings, supported (on the ground) by plywood 'feet'. The plywood and pipes were painted white to minimize solar heating. The towers are stabilized by 8 guy lines (each ¼ inch nylon rope) each tied to 'deadman'

anchors – roughly 40 cm long pieces of bamboo which were buried about 40 cm under the ice. The plywood 'feet' were covered in snow for additional support.

The prototype was deployed in Moore's Bay, from Dec.11-Dec. 21, 2009, where it will collect data for approximately one year. During the installation of the station, we also collected data on various performance metrics.

## Ice

The ice shelf is a key part of our detector. The prototype was deployed at GPS coordinates $78^0$ 44' 523" S, $165^0$ 02' 414" East, about 110 kilometers south of McMurdo station). This is about 1 km from the original site coordinates. There, the ice is about 572 m thick (the details of our measurement are discussed below). The shelf is solid ice at depths greater than about 75 m, but above this it is firn, a gradual transition from packed snow (at the surface) to ice. A plot of the measured density vs. depth for the Ross Ice Shelf is given in Fig. 2 of Ref. [14]; the transition to solid ice occurs at a shallower depth than in central Antarctica.

The surface is flat and relatively featureless. However, from first-hand observations, it is clear that the surface density of the snow varies. We made 4 measurements of snow density; one at the surface, two at a depth of 30-cm, and another about 46 cm deep. For each, we used a saw and a shovel to cut a roughly 10 cm by 10 cm by 15 cm cube of snow and weigh it. These measured snow density at the surface was 0.32 $g/cm^3$, while the other measures were about 0.4 $g/cm^3$. These values are consistent with the roughly 0.36 $g/cm^3$ in Ref. [14].

We also probed the deeper ice by bouncing signals off the ice-water interface; these measurements are discussed below.

## Antennas

The ARIANNA prototype uses four Creative Design Corp. CLP5130-2 [15], 17-element log-periodic dipole antennas, much like VHF/UHF TV antennas. Ideally, the antennas would be oriented pointing down, forming a square, so that two antennas are sensitive to each polarization. Because of the need to avoid the 'deadman' anchors for the guy wires, the antenna pairs (North and South, East and West) were deployed parallel to each other, but not in a square; this departure from the ideal is only relevant in considering relative arrival times for perpendicular antennas (i.e. East and North). The antennas are connected to the prototype box by 6 meters of LMR-600 cable.

The antennas are designed for frequencies (in air) of 105 to 1300 MHz. They are specified to have an 7-8 dBi forward gain in free space, and half power angles of 60-$70^0$ in the E plane, and 110-$130^0$ in the H plane; this provides a good 'field of view' for neutrino hunting. They have 50 ohm impedance and a quoted VSWR of 2:1 or better across our frequency range of interest.

The antennas, which have a 1.4 m boom length, were buried by digging pits about 1.8 m deep, 1.8 m long and 30 cm wide. The antennas were buried pointing down, with their topmost element between 15 cm and 25 cm below the surface. At this depth, the snow-air interface can still affect the received radiation.

Snow has an index of refraction different from air, so the antenna environment will affect both the antenna frequency response and impedance. For radio waves, the index of refraction depends linearly on snow density, and is almost independent of frequency [16]; a density of 0.4 $g/cm^3$ corresponds to a dielectric constant of 1.8 (index of refraction, $n$, about 1.34). This may alter the impedance of the antenna spine, reducing the impedance matching with the preamplifier.

We studied this by comparing the voltage standing wave ratio (VSWR) with the antenna in air (about 1 ½ m above the snow surface), lying flat on the snow, buried in an air-filled pit, and buried in snow. It should be noted that the snow in the refilled hole might not have been the same average density as the undisturbed snow, and, despite efforts to pulverize the snow before filling the hole, that the snow density was not perfectly uniform. Also, the antennas burials were shallow enough that there may be effects from the snow-air interface.

The VSWR was measured with an Agilent "FieldFox" N9912A network analyzer. Figure 2 shows the VSWR for one antenna at the different stages of deployment. There are clear changes in the VSWR, with the positions of various small resonances changing with the conditions. However, in the region from 200 to 1200 MHz, the VSWR is always less than 2.5. At higher frequencies, the VSWR rises, except for the study with the buried antenna; it may be that the increased dielectric constant shifted this increase to higher frequencies.

At low frequencies, the VSWR for the first three conditions increases dramatically below 100 MHz; for the buried antenna, the increase is at a lower frequency, 80 MHz. This is not surprising; for a given frequency, the wavelength in ice is reduced by a factor $1/n$, so one might expect the response of a fixed-size antenna to be shifted to somewhat lower frequencies.

Also, the snow moderates the increase in VSWR seen at low frequencies. The antennas in the air filled holes exhibits significant variation with frequencies; this may be due to some sort of a resonant effect from the width of the hole. Small antenna-to-antenna differences were seen in the VSWR, likely due to slight mechanical differences in assembly.

## Low Noise Amplifiers

The Low Noise Amplifiers (LNAs) amplify the received antennas signals before they are fed into the trigger and the digitizer. One LNA is used per antenna. In the prototype, an 800MHz low pass filter and a 50MHz high pass filter at the input of the LNA block out of band frequencies. This is to remove local generated signals, for example from the 802.11 WiFi link and the Iridium modem and external signals like short wave radio stations. In addition the 50MHz high pass reduces low frequency reflections, which could form standing waves in the antenna cable.

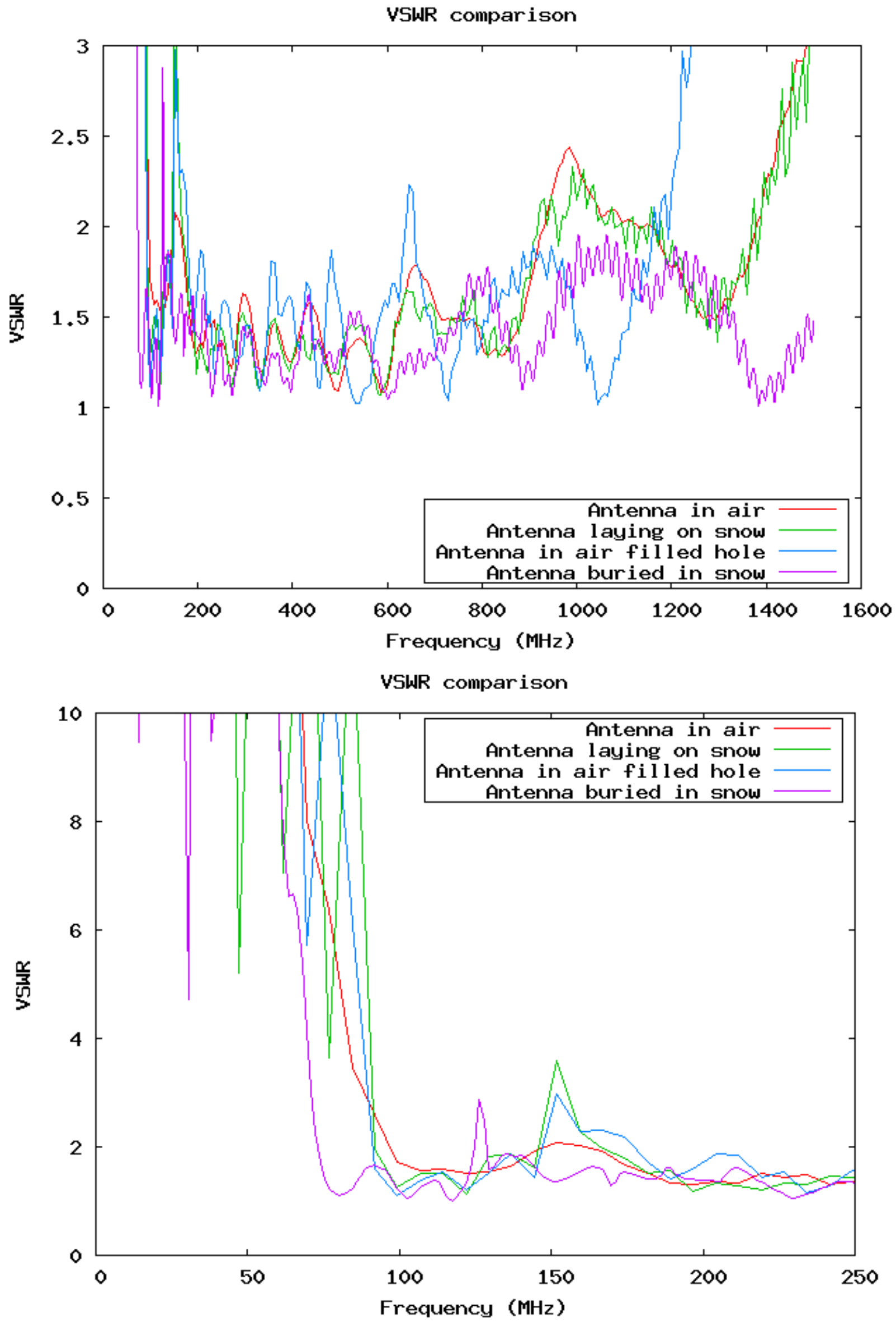

**Figure 3.** Voltage Standing Wave Ratios (VSWR) for one of the log periodic dipole antennas under different conditions, (top) from 50 to 1500 MHz and (bottom) up to 250 MHz. Although the visible peaks and valleys move around depending on the antenna's environment, the VSWR remains below 2.5 in the range of 200 MHz to 1200 MHz, and is generally below 2. It is likely that some of the larger peaks seen in the air filled holes are due to resonances. The VSWR increases dramatically below 100 MHz for the first three cases, and above 80 MHz for the antenna buried in snow.

Each LNA has 4 stages, each consisting of an Avago MGA-68563 GaAs MMIC amplifier. They are broadband, with a gain of 58 dB from 50MHz to 1GHz, and a quoted noise figure of 1.1 dB. The chips are run off of +5V and the power consumption is 250 mW/channel.

To prevent coupling and feedback, each amplifier is individually shielded. In addition, the 4 LNAs are mounted in a shielded box with filtered power feed throughs to prevent possible coupling from other system components like the CPU.

## Data Acquisition and Trigger Circuitry

The data acquisition system is a modified version of the system used on the ANITA balloon flight [5]. Data is recorded with a LABRADOR [17] ASIC which is a switched capacitor array (SCA) waveform sampler. Each channel of the trigger uses a tunnel-diode detector and FPGA discriminator [18]. The whole setup is controlled by a Xilinx Spartan 3 field-programmable gate array (FPGA).

The LABRADOR SCA is a single chip with 8 channels (plus a 9$^{th}$ reference timing channel), each containing 260 (256+4) capacitors. The chip has an analog bandwidth of approximately 1 GHz, and, in ARIANNA, samples at 2.5 Giga-samples/second (GSPS). Each antenna is connected to one channel. A 5$^{th}$ channel is connected to a 40 MHz clock; this is used to calibrate the sampling rate. The other four channels are unused. The chip has 2340 Wilkinson ADCs, so can digitize all of the stored samples in parallel, to 12 bits, in under 50 μs.

The trigger circuit divides the input signal into two frequency bands: 130 to 460 MHz and 650 to 990 MHz. The low-band frequencies are defined by a Mini-Circuits LFCN-320 low-pass filter and a discrete LC circuit, while the high-band is defined by a Mini-Circuits HFCN-650 high-pass filter and a LFCN-800 low-pass filter. The filters have a fairly gradual roll-off; the low-band had a -3dB roll off of 460 MHz, so the intermediate ‘gap’ was not so important. The two bands were used to allow for improved background rejection of low or high frequency noise. Each band feeds a tunnel diode based trigger which acts as a square-law detector. After amplification the tunnel diode feeds an FPGA-based discriminator with a programmable threshold [19].

The circuit threshold is electrically adjustable. Each of the 8 trigger bits (two frequency bands for four channels), are connected to the FPGA, which forms a logical trigger. For most of the prototype running, we used a trigger that ORed the outputs of the two frequency bands from a single antenna, and required at least two of the antennas to trigger.

The prototype station detected significant noise at 300 MHz and 600 MHz, most likely the subharmonics of the 2.4 GHz wireless carrier used to communicate with McMurdo station. So, the thresholds for the lower frequency tunnel diodes were set quite high; most of the triggers were formed using the higher frequency bands.

When the system triggers, the FPGA initiates a LABRADOR digitization cycle and reads out the chip. Data is stored uncompressed, and can be transmitted North over the wireless or Iridium modem.

## Control, Communication, and Housekeeping

The entire system is controlled by a PC104plus based computer [20]. The processor is a 133MHz AMD ELAN, running Slackware 12.2 Linux. The disk is an 8 GByte solid state disk divided into a system and a data partition. The interface to the data acquisition and trigger circuitry is done via USB2.0. In addition to the USB interface the CPU uses several serial ports to talk to peripherals and a Analog/Digital I/O card. The serial ports are assigned to a serial console for debugging work, to an Iridium data modem, and to three GPS receivers.

Only 1 GPS receiver is required, but several different devices were deployed for testing. The first GPS receiver, a Trimble Resolution T, was used before and was known to work. The other two GPS receivers are "M12M Timing Oncore™", which are specified with higher accuracy and lower power consumption. One has an active antenna and the other uses a passive antenna.

The Analog/Digital I/O card is a Diamond – MM-AT-104. It monitors the battery and power supply voltages, measure current draw, and monitors 3 temperature sensors and an anemometer that measures the wind speed. The digital outputs control solid-state power switches which can shut down parts of the station to reduce power consumption. The switches control the RF amplifier, SCA and readout, GPS units, Iridium modem, Trimble GPS, anemometer, heartbeat pulser, and the Ethernet media converters.

Until early February, the system communicated primarily via a wired Ethernet connection to a wireless relay station (erected by the McMurdo Station IT Communications Department) about 15 meters away; the electronics for this relay were removed in early February, before the Austral winter began.

To minimize the possibility of radio interference on the Ethernet connection, a pair of media converters was used. The twisted pair signals from the computer were converted to coaxial cable Ethernet inside the box. Outside the box, a separate unit converted the coaxial signals back to twisted pair Ethernet.

After the wireless link was removed, the station communicated via an Iridium satellite modem. It is configured to 'call home' once every 15 minutes during the summer and 3h during the winter, giving housekeeping data and a few sample waveforms. During this period, most of the data is stored on the flash drive for retrieval next Astral summer.

## Power System

Power is a problem for any Antarctic experiment, especially one that will work through the sunless winter. When fully operational, the deployed ARIANNA prototype draws about 25 Watts; much of this goes to the computer (~7W), the data acquisition and trigger circuitry (~5.5W) and the Ethernet media converters (8W); in the next version of the station, the power consumption will be significantly reduced.

The system has two power sources: four 30-Watt solar panels and a Forgen 1000LT wind driven generator. The power generated is buffered by a 100 Ah gel sealed lead-acid battery to bridge windless periods during the Antarctic night. The battery is about 30% efficient at Antarctic temperatures (buried, the battery temperature is expected to be around -28$^0$C).

The four solar panels provide ample energy to power the system during the summer, and even during ‘shoulder periods’ when the sun sets for part of the day. This will be the primary data collection period for the prototype.

The wind generator produces about 10 Watts in a 15 m/s (~30 knots) wind. This may provide enough power to provide partial operation during the winter with the data acquisition system turned off, so the system is largely recording housekeeping data. A major goal of this mode is to gather environmental data to determine if there is enough wind to generate useful power during the winter.

In case wind power is inadequate, we installed two autonomous temperature loggers, one inside the buried electronics box and one on the frame near the GPS antennas. These loggers record the temperature every 30 minutes for a year, function down to -78$^0$C, and do not require an external power supply.

## Calibrations and the “Heartbeat” system

To verify system performance and overall functionality during the winter, the prototype includes a “Heartbeat” pulser, an Avtech: AVP-AV-1S-P-UCIA, connected through about 10 meters of cable to another CLP5130-2 LPDA antennas, also buried pointing downward, at about a 45 degree angle to the other antennas. The pulser produces a single pulse with a width of about 1.5 ns FWHM and amplitude 6 volts (into 50 ohms). It is programmed to pulse at the beginning of each run, and should provide a quick ‘liveness’ test and also a continuing calibration signal.

## Ice Soundings

During the field season, we studied the properties of the ice sheet by transmitting signals from one buried antenna to another, bouncing them off of the ice/water interface. The Avtech “Heartbeat” pulser was used as a transmitter.

For this, we instrumented the receiving antenna with a Miteq AM-1660-11326 low-noise amplifier; this module has a 58 db gain from 1 kHz to 1 GHz, with a roughly 1.4 dB noise figure. A 1 GHz low pass filter and a 150 MHz high-pass filter were installed before the Miteq LNA; the low pass filter was needed to remove 2.4 GHz RF from a wireless network connection installed in camp, while the 150 MHz filter removed low-frequency oscillations due to reflections in the antenna cable from the antenna and the amplifier.

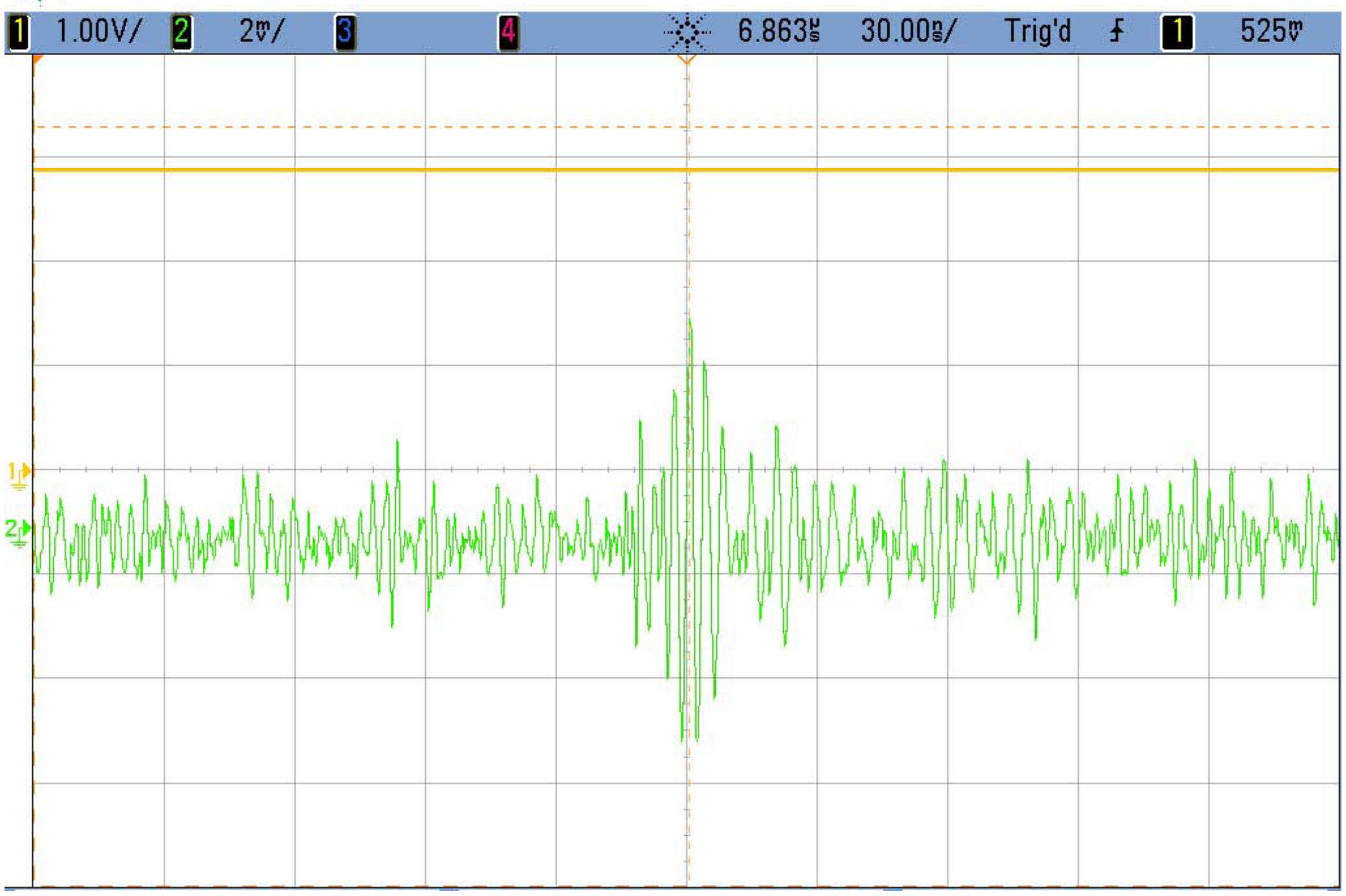

**Figure 4.** Oscilloscope trace of the signal reflected from the ice-water interface. Signal averaging (65536 averages) was used. The main reflected pulse train is about 15 ns long.

The data was recorded by an Agilent DSO7054A oscilloscope with an adjustable delay. The oscilloscope was triggered with the same signal that triggered the Avtech pulser. Data was taken with both parallel antennas orientations (e. g. North to South) and perpendicular (e.g. North to East). Figure 4 shows an example of a reflected trace, using signal averaging; the pulse-to-pulse jitter was small. The reflected pulse is about 20 ns long. This is consistent with what is expected from a smooth ice-water interface. For comparison, Fig. 5 shows a similar signal, taken when two antennas were placed facing each other (nose-to-nose) with a 2.7 m spacing, in air. Here, no low-pass filter was used. The signal is somewhat longer for the air-to-air transmission, likely because there was less attenuation of low-frequency components. However, the two signals are quite compatible, indicating that the ice-water interface provides smooth reflection

For both the parallel and perpendicular orientations, a return signal was observed at time $t$ = 6.745 μs after the original downgoing pulse; the pulse amplitude at the oscilloscope input was about 6 mV peak-to-peak for the parallel orientation, and about 1/3 that for the perpendicular orientation. This is consistent with the expectations due to the antenna radiation patterns. Signal averaging was required to accurately measure these signals. We attribute a ± 15 ns uncertainty to the round-trip travel time, to account for uncertainties in the cable lengths, geometry, and arrival times (due to the pulse width).

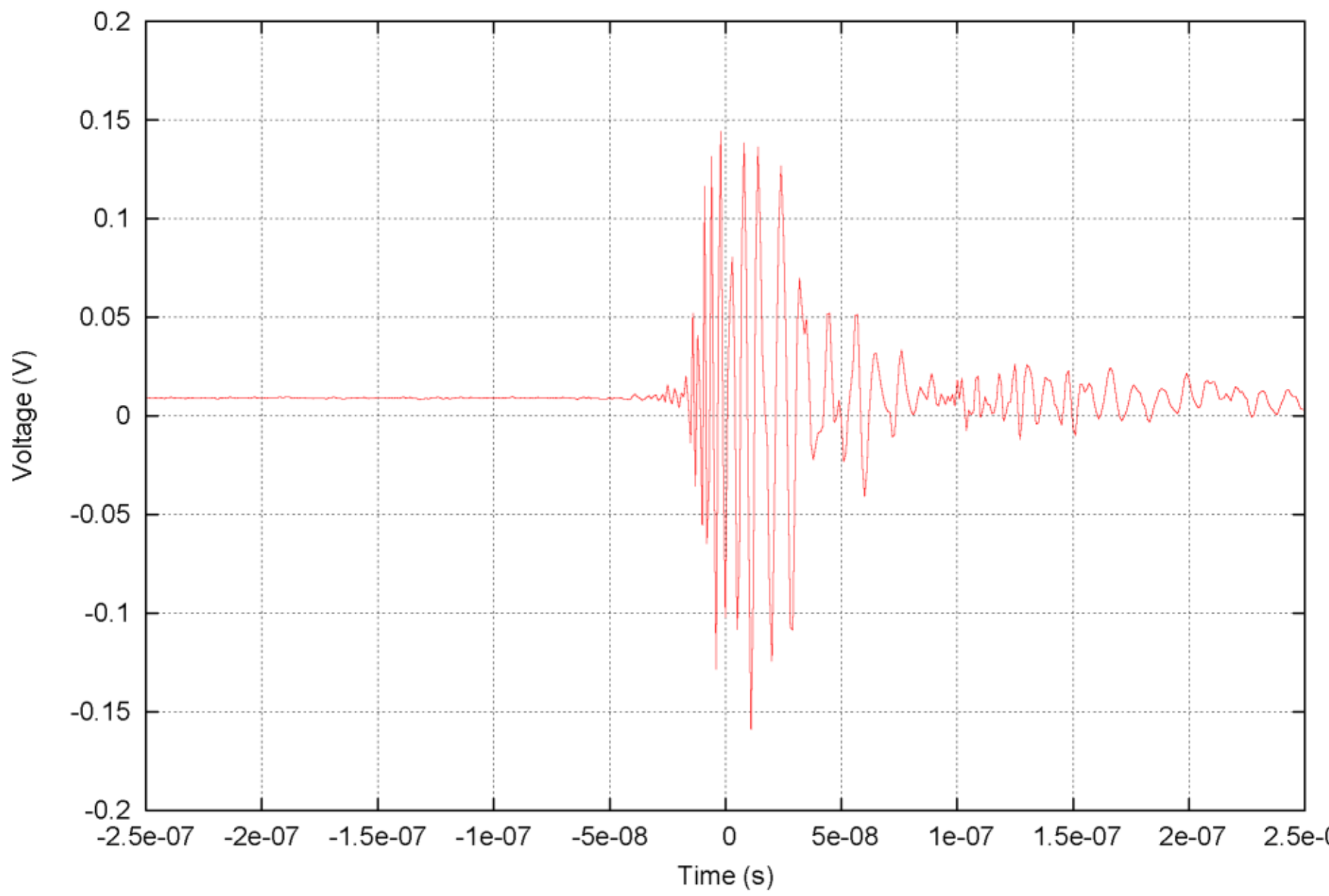

**Figure 5.** The transmitted signal from an Avtech pulser sent between two CLP5130-2 LPDA antennas sitting head-to-head, in air, separated by 2.7 m.

The relationship between the round-trip travel time and the ice thickness depends on the index of refraction in the ice, which itself depends on the density, and, to a much lesser extent, on the possible presence of impurities [14, 21]. The ice in Moore's bay is glacial ice, so should be free of impurities; we do not see any evidence of infiltrating brine layers in the radio reflections.

Our calculation follows Ref. [14]. The index of refraction, *n,* is the square root of the dielectric constant ε. The index of refraction of pure, solid ice has a very slight temperature dependence, $\varepsilon = (3.18\pm0.01) + (8\times10^{-4}\ T)$, where T is in Centigrade. The ice temperature varies with depth, from near $0^0$C at the ice-water interface, to about $-20^0$C just below the surface [22]. We assume a mean temperature (averaged over the radio path) of $-10 \pm 5^0$C, giving an average index of refraction of n=1.78 for solid ice. The upper 75 m of ice is firn, which has a lower density, where the radio signal will travel faster; ($n$-1) scales linearly with the density. Dowdeswell and Evans [22] treat this region by including a correction in the calculated ice thickness:

$$D = \frac{1}{2}\left(\frac{ct}{n} + z_f\right)$$

where c is the speed of light. For the Ross Ice Shelf, the correction $z_f$ is +7±2 m. From this, we find a thickness of 572 m, with uncertainties due to the travel time (5 m), temperature (1 m), index of refraction (1 m) and firn correction (2 m), giving a total uncertainty of ±6 m.

## Conclusions

ARIANNA is designed to detect cosmic GZK neutrinos; it will eventually comprise an array with an active volume of at least 100 $km^3$ with a threshold of around $10^{17}$ eV.

We have built a prototype detector which reads out four log-periodic dipole antennas, using switched capacitor array digitizers. The prototype was deployed in Moore's Bay, Antarctica, in December 2009, where it will collect data for one year, providing valuable information about conditions during the Antarctic winter and backgrounds for a neutrino search. We also bounced radio signals off the ice-water interface, and measured an ice thickness of $572 \pm 6$ m.

Work on the next prototype is beginning now. This will be an integrated design, with simpler electronics interfaces and greatly reduced power consumption. It will likely incorporate 8 antennas per station, quite possibly including one upward-facing antenna for background rejection. This design will be used in a 7 station array to be deployed within 2-3 years.

## Acknowledgements

We thank Martha Story for invaluable help with camp management and David Saltzberg for useful conversations. Gary Varner and Larry Ruckman developed much of the data acquisition system. We thank the National Science Foundation, Office of Polar Programs for major logistical support. This work was funded in part by the National Science Foundation under grant number 0653266, by the Department of Energy under contract number DE-AC-76SF-00098 and by the Royal Society.